\documentclass[amsmath, pra, twocolumn, showpacs, floatfix]{revtex4-1}
\usepackage{graphicx}
\usepackage{dsfont}

\begin{document}   

\title{Torque of guided light on an atom near an optical nanofiber}
 
\author{Fam Le Kien}
\affiliation{Quantum Systems Unit, Okinawa Institute of Science and Technology Graduate University, Onna, Okinawa 904-0495, Japan}

\author{Thomas Busch}
\affiliation{Quantum Systems Unit, Okinawa Institute of Science and Technology Graduate University, Onna, Okinawa 904-0495, Japan}

\date{\today}

\begin{abstract}
We calculate analytically and numerically the axial orbital and spin torques of guided light on a two-level atom near an optical nanofiber. We show that the generation of these torques is governed by the angular momentum conservation law in the Minkowski formulation. The orbital torque on the atom near the fiber has a contribution from the average recoil of spontaneously emitted photons.  Photon angular momentum and atomic spin angular momentum can be converted into atomic orbital angular momentum.  The orbital and spin angular momenta of the guided field are not transferred separately to the orbital and spin angular momenta of the atom.
\end{abstract}

\maketitle

\section{Introduction}

The ability to transfer momentum between light and atoms, molecules, or material particles is one of the cornerstones of light-matter interactions and has found many applications in physics and technology \cite{CCT:07,Roadmap:17}.
For the electromagnetic field in a dielectric medium, several formulations for the linear and angular momentum densities can be found in the literature \cite{Brevic1979,Pfeifer2007,theme issue,Bliokh2017a,Bliokh2017b}.
For example, at the simplest level, the momentum of a single plane-wave photon is $p_{\mathrm{A}}=\hbar k_0/n$ in the Abraham formulation and $p_{\mathrm{M}}=\hbar k_0 n$ in the Minkowski formulation, where $k_0$ is wave number in free space and $n$ is the refractive index of the medium. The difference between these formulations originates from the fact that, in the different existing theories,  the divisions of the total energy-momentum tensor into electromagnetic and matter contributions are different from each other and depend on the choice of observable \cite{Pfeifer2007}. Depending on specific situations, one of the forms of momentum appears as the natural, experimentally observed momentum  \cite{Brevic1979,Pfeifer2007,theme issue}. It has been shown for a single atom interacting with a light pulse that both the Abraham and Minkowski momenta of photons have identifiable roles associated, respectively, with the kinetic and canonical momenta of the atom \cite{Hinds2009}. The mass-polariton theory of light in a medium gives an unambiguous physical meaning to the Abraham and Minkowski momenta \cite{Partanen2017a}.
According to this theory, a light pulse propagating in a medium is made of mass-polariton quasiparticles, which are quantized coupled states of the field and an atomic mass density wave, driven forward  by the optical force. The total momentum of a mass-polariton quasiparticle is the Minkowski momentum, the contribution from the field is the Abraham momentum, and the difference is carried by the mass density wave \cite{Partanen2017a,Partanen2017b,Partanen2018a,Partanen2018b,Partanen2019}.

Angular momentum transfer from light to matter has been examined in a large number of systems in recent years \cite{theme issue,Franke-Arnold2017}. These include the transfers of angular momentum of a paraxial light beam to particles \cite{Dunlop1995,Dunlop1996,Simpson1997,Dholakia2003}, 
atoms \cite{Babiker1994,Picon2010,Afanasev2013,Lembessis2013}, 
molecules \cite{Babiker2002,Veenendaal2007},
ensembles of cold atoms \cite{Inoue2006,Moretti2009,Nicolas2014,Radwell2015},
and Bose-Einstein condensates \cite{Madison2000,Andersen2006,Ryu2007,Wright2008}. 
Periodic exchange of angular momentum between an atom and a reflecting surface has been studied  \cite{Donaire2015}. The optical torque on a two-level system near a strongly nonreciprocal medium has been calculated \cite{Monticone2018}.

One prominent and experimentally relevant example where the differences between the Abraham and Minkowski formulations of linear and angular momenta are important is the case of guided light of nanofibers.
Such fibers are vacuum-clad ultrathin optical fibers that allow tightly radially confined light to propagate over a long distance and to interact efficiently with nearby atoms \cite{TongNat03,review2016,review2017,Nayak2018}. Due to the cylindrical symmetry of the nanofibers, light in a higher-order mode may have a large angular momentum. 
The cross-sectional profile of the guided light of a vacuum-clad nanofiber has two parts, 
the inner part, which is confined in the dielectric medium of the fiber,
and the outer part, which extends radially in the vacuum outside the fiber boundary \cite{Tong04,fibermode}. 
Therefore, the magnitudes of linear and angular momenta of guided light depend on the formulations for these characteristics \cite{Partanen2018a,Fam2006}. Calculating the Abraham angular momentum of a photon in a quasicircularly polarized guided mode with an azimuthal mode order $l$ shows that it increases with increasing $l$  \cite{highorder}, but is different from $l\hbar$ \cite{Partanen2018a,Fam2006,highorder}. Conversely,  the Minkowski angular momentum per photon is quantized to be exactly equal to $l\hbar$ \cite{Partanen2018a,Fam2006,Bliokh2018}. It is desirable to know what form is appropriate for transfers of angular momentum from guided light to atoms. Since light in a high-order guided mode can have a large angular momentum, such transfers can be used to generate, control, and manipulate the orbital and spin angular momenta of atoms, molecules, and particles. 

When considering the optical forces and torques that stem from the transfers of linear and angular momenta from guided light to atoms,  the near-field nature of the guided field requires careful treatment. In particular,  for atoms near a nanofiber, spontaneous emission can become asymmetric with respect to opposite propagation directions \cite{Fam2014,Petersen2014,Mitsch14b,sponhigh}  due to spin-orbit coupling of light carrying transverse spin angular momentum  \cite{Zeldovich,Zeldovich1,Zeldovich2,Bliokh review,Bliokh2014,Bliokh review2015,Bliokh2015,Banzer review2015,Lodahl2017}. 
Asymmetric spontaneous emission may lead to a nonzero average spontaneous emission recoil and, hence, may contribute to the optical force on the atoms. In particular, a lateral spontaneous emission recoil force may arise for an initially excited atom near a nanofiber \cite{Scheel2015,Jacob2016}.
Recently, the full vector structure of the force of guided light on an atom near a vacuum-clad ultrathin optical fiber has been investigated \cite{chiralforce,NJP}. It is clear that the azimuthal component of the force leads to an axial torque and consequently to a transfer of angular momentum from guided light to the orbital motion of the atom.
 
In this work, we study the transfer of angular momentum from a guided driving light field to a two-level atom near a vacuum-clad ultrathin optical fiber. We calculate the torque on the atom by employing
the previous results of  \cite{chiralforce,NJP} for the force, which were obtained by using the Hamiltonian formalism and the mode function expansion technique.
We show that the generation of the axial orbital and spin torques of guided light is governed by the angular momentum conservation law with the photon angular momentum in the Minkowski formulation.
We find that the orbital torque on the atom near the fiber has a contribution  from the averaged recoil of spontaneously emitted photons. We show that the orbital and spin angular momenta of the guided field are not transferred separately to the orbital and spin angular momenta of the atom. 
 
The paper is organized as follows. In Sec.~\ref{sec:model} we describe the model system and briefly review the force of guided light on an atom.
In Sec.~\ref{sec:torque} we study the axial orbital and spin torques of guided light on the atom. 
Our conclusions are given in Sec.~\ref{sec:summary}.

\section{Model system and force of guided light on an atom}
\label{sec:model}

In this section, we describe the model system and briefly review
the previous results of  \cite{chiralforce,NJP} for the force of guided light on an atom, which were  obtained by using the Hamiltonian formalism and the mode function expansion technique.

\begin{figure}[htbp]
\centering\includegraphics[width=0.8 \linewidth]{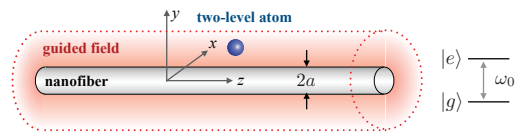}
\caption{Two-level atom driven by guided light of a vacuum-clad ultrathin optical fiber.}
\label{fig1}
\end{figure}

We consider a two-level atom driven by a near-resonant classical guided field with optical frequency $\omega_c$ and envelope $\boldsymbol{\mathcal{E}}$ near a vacuum-clad ultrathin optical fiber (see Fig.~\ref{fig1}). 
The atom has an upper energy level $|e\rangle$ and a lower energy level $|g\rangle$, with energies $\hbar\omega_e$ and $\hbar\omega_g$. The atomic transition frequency is $\omega_0=\omega_e-\omega_g$. The fiber has a cylindrical dielectric core of radius $a$ and refractive index $n_1>1$ and an infinite vacuum cladding of refractive index $n_2=1$. We are interested in vacuum-clad silica-core ultrathin fibers with diameters in the range of hundreds nanometers, which can support only the fundamental HE$_{11}$ mode and a few higher-order modes in the optical region. Such optical fibers are usually called nanofibers \cite{TongNat03,review2016,review2017,Nayak2018}. In view of the very low losses of silica in the wavelength range of interest, we neglect material absorption. We use Cartesian coordinates $\{x,y,z\}$, where $z$ is the coordinate along the fiber axis, and also cylindrical coordinates $\{r,\varphi,z\}$, where $r$ and $\varphi$ are the polar coordinates in the transverse plane $xy$.

The atom interacts with the classical guided driving field $\boldsymbol{\mathcal{E}}$ and the quantum electromagnetic field. In the presence of the fiber, the quantum field  can be decomposed into the contributions from guided and radiation modes \cite{fiber books, fiber books 1, fiber books 2}.  Then, the Hamiltonian for the atom-field interaction in the dipole approximation is \cite{chiralforce,NJP} 
\begin{eqnarray}\label{c13}
H_{\mathrm{int}}&=&-\frac{\hbar}{2}\Omega\sigma_{eg}e^{-i(\omega_c-\omega_0)t}
-i\hbar\sum_{\alpha}G_{\alpha}\sigma_{eg} a_{\alpha}e^{-i(\omega-\omega_0)t}\nonumber\\
&&\mbox{}
-i\hbar\sum_{\alpha}\tilde{G}_{\alpha}\sigma_{ge} a_{\alpha}e^{-i(\omega+\omega_0)t}
+\mbox{H.c.},
\end{eqnarray}
where $\sigma_{ij}=|i\rangle\langle j|$ with $i,j=e,g$ are the atomic operators, $a_{\alpha}$ and $a_{\alpha}^\dagger$ are the photon operators, 
$\Omega=\mathbf{d}\cdot\boldsymbol{\mathcal{E}}/\hbar$ is the Rabi frequency of the driving field,
with $\mathbf{d}=\mathbf{d}_{eg}=\langle e|\mathbf{D}|g\rangle$ being the matrix element of the atomic dipole operator $\mathbf{D}$,
and $G_\alpha$ and $\tilde{G}_\alpha$ are the coupling coefficients for the interaction between the atom and the quantum field  \cite{chiralforce,NJP}.
The notations $\alpha=\mu,\nu$ and $\sum_{\alpha}=\sum_{\mu}+\sum_{\nu}$ stand for the mode index and the mode summation.
The index $\mu=(\omega N f p)$ labels guided modes, where $\omega$ is the mode frequency, $N=\mathrm{HE}_{lm}$, EH$_{lm}$, TE$_{0m}$, or TM$_{0m}$ is the mode type, with $l=1,2,\dots$ and $m=1,2,\dots$ being the azimuthal and radial mode orders, $f=\pm1$ denotes the forward or backward propagation direction along the fiber axis $z$, and $p=\pm1$ for HE and EH modes and $0$ for TE and TM modes is the phase circulation direction index \cite{fiber books, fiber books 1, fiber books 2}. The longitudinal propagation constant $\beta$ of a guided mode is determined by the fiber eigenvalue equation. 
The index $\nu=(\omega \beta l p)$ labels radiation modes, where $\beta$ is the longitudinal propagation constant,  
$l=0,\pm1,\pm2,\dots$ is the mode order, and $p=+,-$ is the mode polarization index.
The longitudinal propagation constant $\beta$ of a radiation mode of frequency $\omega$ can vary continuously, from $-kn_2$ to $kn_2$ (with $k=\omega/c$).
The notations $\sum_{\mu}=\sum_{N fp}\int_0^{\infty}d\omega$ and $\sum_{\nu}=\sum_{lp}\int_0^{\infty}d\omega\int_{-kn_2}^{kn_2}d\beta$ denote the summations over guided and radiation modes.
We emphasize that the atom can absorb a photon in the classical guided driving field and then emit a photon into the quantum guided and radiation modes. This is the reason why we need to include the quantum guided and radiation modes in our model. 

The expressions for the coupling coefficients $G_{\alpha}$ and $\tilde{G}_{\alpha}$ with $\alpha=\mu,\nu$ are given by   
\begin{eqnarray}\label{c14}
G_{\mu}&=&\sqrt{\frac{\omega\beta'}{4\pi\epsilon_0\hbar}}\;
(\mathbf{d}\cdot\mathbf{e}^{(\mu)})e^{i(f\beta z+pl\varphi)},\nonumber\\
G_{\nu}&=&\sqrt{\frac{\omega}{4\pi\epsilon_0\hbar}}\;
(\mathbf{d}\cdot\mathbf{e}^{(\nu)})e^{i(\beta z+l\varphi)},
\end{eqnarray}
and
\begin{equation}\label{c15}
\begin{split}
\tilde{G}_{\mu}&=\sqrt{\frac{\omega\beta'}{4\pi\hbar\epsilon_0}}\;
(\mathbf{d}^*\cdot\mathbf{e}^{(\mu)})e^{i(f\beta z+pl\varphi)},\\
\tilde{G}_{\nu}&=\sqrt{\frac{\omega}{4\pi\hbar\epsilon_0}}\;
(\mathbf{d}^*\cdot\mathbf{e}^{(\nu)})e^{i(\beta z+l\varphi)},
\end{split}
\end{equation}
where $\mathbf{e}^{(\mu)}$ and $\mathbf{e}^{(\nu)}$ are the normalized mode functions given in \cite{fiber books, fiber books 1, fiber books 2,sponhigh}.
An important property of the mode functions of the hybrid HE and EH modes and the TM modes is that the longitudinal
component $e_z^{(\mu)}$ is nonvanishing and in quadrature with the radial component $e_r^{(\mu)}$.
We note that in the Hamiltonian $H_{\mathrm{int}}$ given by Eq. (\ref{c13}) we have used the rotating-wave approximation for the driving field but not for the quantum field.

In a semiclassical treatment, the center-of-mass motion of the atom is governed by the force 
$\mathbf{F}= -\langle\boldsymbol{\nabla} H_{\mathrm{int}}\rangle$ \cite{coolingbook,dipole force,dipole force 1,dipole force 2}.
According to \cite{chiralforce,NJP}, we have
\begin{equation}\label{c25}
\mathbf{F}=\mathbf{F}^{\mathrm{(drv)}}+\rho_{ee}\mathbf{F}^{\mathrm{(spon)}}+\rho_{ee}\mathbf{F}^{(\mathrm{vdW})e}
+\rho_{gg}\mathbf{F}^{(\mathrm{vdW})g},             
\end{equation}
where
\begin{equation}\label{c26}
\mathbf{F}^{\mathrm{(drv)}}=\frac{\hbar}{2}(\rho_{ge}\boldsymbol{\nabla}\Omega+\rho_{eg}\boldsymbol{\nabla}\Omega^*)             
\end{equation}
is the force produced by the interaction with the driving field,
\begin{equation}\label{c27}
\mathbf{F}^{\mathrm{(spon)}}=i\pi\hbar\sum_{\alpha_0}(G_{\alpha_0}^*\boldsymbol{\nabla}G_{\alpha_0}-G_{\alpha_0}\boldsymbol{\nabla}G_{\alpha_0}^*)    \end{equation}
is the force resulting from the average recoil of spontaneous emission of the atom in the excited state \cite{Scheel2015}, and
\begin{equation}\label{c28}
\mathbf{F}^{(\mathrm{vdW})e}=\hbar\boldsymbol{\nabla}\mathcal{P}\sum_{\alpha}\frac{|G_{\alpha}|^2}{\omega-\omega_0}
\end{equation}
and
\begin{equation}\label{c29}
\mathbf{F}^{(\mathrm{vdW})g}=\hbar\boldsymbol{\nabla}\mathcal{P}\sum_{\alpha}\frac{|G_{\alpha}|^2}{\omega+\omega_0}    \end{equation}
are the forces resulting from the van der Waals potentials for the excited and ground states, respectively.
We have also introduced the notations $\rho_{ij}=\langle i|\rho|j\rangle$ with $i,j=e,g$ for the matrix elements of the density operator $\rho$ for the atomic internal state.
In Eq.~(\ref{c27}), the notation $\alpha_0$ is the label of a resonant guided mode $\mu_0=(\omega_0 N f p)$ or a resonant radiation mode $\nu_0=(\omega_0 \beta l p)$.
We note that $\mathbf{F}^{(\mathrm{scatt})}\equiv\rho_{ee}\mathbf{F}^{\mathrm{(spon)}}$ is the force produced by the average recoil of the photons that are scattered from the atom with the excited-state population $\rho_{ee}$.

The forces $\mathbf{F}^{(\mathrm{vdW})e}$ and $\mathbf{F}^{(\mathrm{vdW})g}$ are produced by the van der Waals potentials $U_e$ and $U_g$ \cite{Buhmann2004}, that is, $\mathbf{F}^{(\mathrm{vdW})e}=-\boldsymbol{\nabla} U_e$ and $\mathbf{F}^{(\mathrm{vdW})g}=-\boldsymbol{\nabla} U_g$. 
These body-induced potentials are given as
\begin{eqnarray}\label{c30}
U_e&=&-\hbar\mathcal{P}\sum_{\alpha}\frac{|G_{\alpha}|^2}{\omega-\omega_0}-\delta E_e^{(\mathrm{vac})},
\nonumber\\
U_g&=&-\hbar\mathcal{P}\sum_{\alpha}\frac{|G_{\alpha}|^2}{\omega+\omega_0}-\delta E_g^{(\mathrm{vac})},  
\end{eqnarray}
where $\delta E_e^{(\mathrm{vac})}$ and $\delta E_g^{(\mathrm{vac})}$ are the energy level shifts induced by the vacuum field in free space (in the absence of the fiber).
Note that $\delta E_e^{(\mathrm{vac})}-\delta E_g^{(\mathrm{vac})}=\hbar\delta\omega_0^{(\mathrm{vac})}$, where $\delta\omega_0^{(\mathrm{vac})}$ is the Lamb shift of the transition frequency of the atom in free space. 
The detuning of the field from the atom near the fiber can be written as $\Delta=\Delta_0-(U_e-U_g)/\hbar$, where $\Delta_0=\omega_L-\omega_0-\delta\omega_0^{(\mathrm{vac})}$ is the detuning of the field from the atom in the absence of the fiber.

In the case of atoms in free space, spontaneous emission is symmetric with respect to opposite propagation directions. In this case, we have $\mathbf{F}^{\mathrm{(spon)}}=0$.
However, in the case of atoms near an object, spontaneous emission may become asymmetric with respect to opposite propagation directions  \cite{Fam2014,Petersen2014,Mitsch14b,sponhigh}. This directional effect is due to spin-orbit coupling of light carrying transverse spin angular momentum  \cite{Zeldovich,Zeldovich1,Zeldovich2,Bliokh review,Bliokh2014,Bliokh review2015,Bliokh2015,Banzer review2015,Lodahl2017}. Asymmetric spontaneous emission may lead to a nonzero average spontaneous emission recoil and, hence, may contribute to the optical force on the atoms. In particular, an axial lateral spontaneous emission recoil force $F_z^{(\mathrm{spon})}$ may arise for an initially excited atom near a nanofiber \cite{Scheel2015,Jacob2016}. 

\section{Axial orbital and spin torques of the guided light on the atom}
\label{sec:torque}

The azimuthal force component $F_\varphi$ is responsible for the rotational motion of the atom around the fiber axis. 
The axial component of the orbital torque on the atom is $T_z=rF_\varphi$. 
This torque component characterizes the rate of the change of the axial component of the orbital angular momentum of the atom.
We use Eqs.~\eqref{c25}--\eqref{c29} to calculate the azimuthal force $F_\varphi$ and the axial orbital torque $T_z$. 
Then, we obtain
\begin{equation}\label{c25a}
T_z=T_z^{\mathrm{(drv)}}+\rho_{ee}T_z^{\mathrm{(spon)}}
+\rho_{ee}T_z^{(\mathrm{vdW})e}+\rho_{gg}T_z^{(\mathrm{vdW})g}.
\end{equation}
The term $T_z^{\mathrm{(drv)}}$ is the axial torque component produced by the driving field and is given as
\begin{equation}\label{c26a}
T_z^{\mathrm{(drv)}}=\frac{\hbar}{2}\left(\rho_{ge}\frac{\partial\Omega}{\partial \varphi}
+\rho_{eg}\frac{\partial\Omega^*}{\partial \varphi}\right).
\end{equation}
The term $T_z^{\mathrm{(spon)}}$ is the axial torque component produced by the average recoil of spontaneous emission of the atom in the excited state $|e\rangle$
and is given as 
\begin{equation}\label{c27a}
T_z^{\mathrm{(spon)}}=
i\pi\hbar\sum_{\alpha_0}\left(G_{\alpha_0}^*\frac{\partial G_{\alpha_0}}{\partial \varphi}
-G_{\alpha_0}\frac{\partial G_{\alpha_0}^*}{\partial \varphi}\right).
\end{equation}
Note that $T_z^{(\mathrm{scatt})}\equiv\rho_{ee}T_z^{\mathrm{(spon)}}$ is the axial torque component produced by the average recoil of the photons that are scattered from the atom with the excited-state population $\rho_{ee}$.
The terms 
\begin{eqnarray}
T_z^{(\mathrm{vdW})e}&=&-\frac{\partial U_e}{\partial\varphi},\nonumber\\
T_z^{(\mathrm{vdW})g}&=&-\frac{\partial U_g}{\partial\varphi}
\end{eqnarray}   
are the axial torques resulting from the van der Waals potentials $U_e$ and $U_g$ for the excited and ground states.

We emphasize again that, in the case of atoms in free space, spontaneous emission is symmetric with respect to opposite propagation directions. In this case, we have $T_z^{\mathrm{(spon)}}=0$, that is, the recoil of spontaneously emitted photons in average does not contribute to the axial orbital torque $T_z$ on the atom. Hence, $T_z^{\mathrm{(spon)}}$ was not considered in the previous work on the angular momentum transfer \cite{Babiker1994,Picon2010,Afanasev2013}.  
However, in the case of atoms near an object, we may have $T_z^{\mathrm{(spon)}}\not=0$.

In the following we calculate the components of the axial orbital torque $T_z$ on the atom near the fiber.
First, we calculate the axial orbital torque $T_z^{\mathrm{(drv)}}$ produced by the interaction with the guided driving field. 
We assume that this field is prepared in a quasicircularly hybrid HE or EH mode, a TE mode, or a TM mode.
Such a mode can be labeled by an index $\mu_c=(\omega_c N_c f_c p_c)$,
where $\omega_c$ is the driving field frequency, $N_c=\mathrm{HE}_{l_cm_c}$, EH$_{l_cm_c}$, TE$_{0m_c}$, or TM$_{0m_c}$ is the mode type, $f_c=\pm1$ denotes the forward or backward propagation direction along the fiber axis $z$, and $p_c=\pm1$ for HE and EH modes and $0$ for TE and TM modes is the phase circulation direction index \cite{fiber books, fiber books 1, fiber books 2}.  
Here,  $l_c=1,2,\dots$ for HE and EH modes and 0 for TE and TM modes is the azimuthal mode order, and $m_c=1,2,\dots$ is the  radial mode order.  Then,  the amplitude of the driving field can be written as 
\begin{equation}\label{c9}
\boldsymbol{\mathcal{E}}=\mathcal{A}[\hat{\mathbf{r}}e_r^{(\mu_c)}(r)+\hat{\boldsymbol{\varphi}}e_\varphi^{(\mu_c)}(r)+\hat{\mathbf{z}}e_z^{(\mu_c)}(r)]e^{if_c\beta_c z+ip_cl_c\varphi},
\end{equation}
where $e_{r,\varphi,z}^{(\mu_c)}(r)$ are the cylindrical components of the reduced mode function and depend on $r$ but not on $\varphi$ and $z$,
and $\mathcal{A}$ is a constant determined by the field power. 

We introduce the notations  $V_0=V_z$ and $V_{\pm1}=\mp(V_x\pm iV_y)/\sqrt{2}$
for the spherical tensor components of an arbitrary complex vector $\mathbf{V}$.
We assume that the dipole matrix element vector $\mathbf{d}$ has a single nonzero spherical tensor element $d_q$, where $q=0,\pm1$.
Such a transition can be realized between the magnetic levels $M'$ and $M=M'-q$ of an electric-dipole emission line of an alkali-metal atom.
The corresponding type of the atomic transition is $\pi$ for $q=0$ and $\sigma_{\pm}$ for $q=\pm1$ with respect to the quantization axis $z$.
Then, the Rabi frequency $\Omega$ for the field-atom interaction is 
\begin{equation}\label{b9f}
\Omega=(-1)^q d_{q}\mathcal{E}_{-q}/\hbar.
\end{equation}
It follows from Eq.~\eqref{c9} that
\begin{equation}\label{b9e}
\mathcal{E}_{-q}=\mathcal{A}e_{-q}^{(\mu_c)}(r)e^{-iq\varphi} e^{if_c\beta_c z+ip_cl_c\varphi}.
\end{equation}
This leads to 
\begin{equation}\label{b9b}
\frac{\partial\Omega}{\partial\varphi}=i(p_cl_c-q)\Omega. 
\end{equation}
Then, Eq.~\eqref{c26a} yields
\begin{equation}\label{b9g}
T_z^{\mathrm{(drv)}}=-(p_cl_c-q)\hbar\mathrm{Im}(\rho_{ge}\Omega).
\end{equation}
On the other hand,
the time evolution of the population $\rho_{ee}$ of the atomic upper state is governed by the equation  \cite{coolingbook}
\begin{equation}\label{b9c}
\dot{\rho}_{ee}=-\mathrm{Im}(\Omega\rho_{ge})-\Gamma{\rho}_{ee},
\end{equation}
where 
\begin{equation}\label{b9d}
\Gamma=2\pi\sum_{\alpha_0}|G_{\alpha_0}|^2 
\end{equation}
is the rate of spontaneous emission of the atom in the presence of the fiber \cite{sponhigh}.
Hence, the axial component of the orbital torque resulting from the interaction with the driving field is found to be
\begin{equation}\label{b7}
T_z^{\mathrm{(drv)}}=(p_cl_c-q)\hbar(\Gamma\rho_{ee}+\dot{\rho}_{ee}).
\end{equation}
It is clear that $T_z^{\mathrm{(drv)}}$ is produced by
the force $F_\varphi^{\mathrm{(drv)}}=T_z^{\mathrm{(drv)}}/r=(p_cl_c-q)\hbar(\Gamma\rho_{ee}+\dot{\rho}_{ee})/r$, which is the azimuthal pressure force component.   

Equation \eqref{b7} describes the exchange of angular momentum between the guided driving field and the atom in the excitation process. According to \cite{Partanen2018a,Fam2006,Bliokh2018}, the canonical angular momentum of a photon in the guided driving field in the Minkowski formulation is $p_cl_c\hbar$ (see Appendix \ref{sec:appendix}). The change of the spin angular momentum of the atom due to an upward transition is $q\hbar$. The scattering rate is equal to $\Gamma\rho_{ee}$ \cite{coolingbook}. 
The upward-transition (photon-absorption) rate is equal to $\Gamma\rho_{ee}+\dot{\rho}_{ee}$, which is the sum of the scattering rate $\Gamma\rho_{ee}$
and the atomic excitation rate $\dot{\rho}_{ee}$ \cite{coolingbook}. 
Then, it is clear from Eq.~\eqref{b7} that the angular momentum of absorbed guided photons is converted into the orbital and spin angular momenta of the atom. Thus, Eq.~\eqref{b7} shows that the generation of the torque of light on the atom is governed by the conservation of the total angular momentum of the atom-field system. 
Moreover, Eq.~\eqref{b7} confirms that the photon recoil imparted on an atom near a nanofiber is of the Minkowski form.
This conclusion is consistent with the results for the forces of stationary light fields on atoms in dielectric media or near objects 
\cite{Jones1978,Pritchard2005,Milonni2005,Milonni2010,Anzetta2018a,Anzetta2018b} and also with the results for the linear and angular momenta of mass-polariton quasiparticles formed by guided light pulses in optical fibers \cite{Partanen2018a} or Laguerre-Gaussian light pulses in bulk media \cite{Partanen2018b}. However, it is different from the result for a light pulse interacting with a single atom in free space \cite{Hinds2009}.
Equation \eqref{b7} is a key result of the present paper.

The spin torque of light on an oscillating electric dipole is given by \cite{Bliokh2014,Jackson} 
\begin{equation}\label{b12a}
\mathbf{Q}^{\mathrm{(drv)}}=\frac{1}{2}\mathrm{Re}(\boldsymbol{\mathcal{D}}^*\times\boldsymbol{\mathcal{E}}), 
\end{equation}
where $\boldsymbol{\mathcal{D}}$ is the envelope of the dipole positive-frequency component.
For the two-level atom considered here, we have $\boldsymbol{\mathcal{D}}=2\mathbf{d}^*\rho_{eg}$. 
In the case where the dipole matrix element vector $\mathbf{d}$ has a single nonzero spherical tensor component $d_q$, 
the axial component of the spin torque resulting from the interaction with the driving field is found to be 
\begin{equation}\label{b12b}
Q_z^{\mathrm{(drv)}}=-q\hbar\mathrm{Im}(\Omega\rho_{ge}). 
\end{equation}
Using Eq.~\eqref{b9c}, we obtain
\begin{equation}\label{b12}
Q_z^{\mathrm{(drv)}}=q\hbar(\Gamma\rho_{ee}+\dot{\rho}_{ee}).
\end{equation}
Equation \eqref{b12} indicates that the axial spin torque $Q_z^{\mathrm{(drv)}}$ resulting from the interaction with the driving field is the product of 
the change $q\hbar$ of the spin angular momentum of the atom per upward transition and the photon-absorption rate $\Gamma\rho_{ee}+\dot{\rho}_{ee}$.

We note that 
\begin{equation}\label{b12d}
T_z^{\mathrm{(drv)}}+Q_z^{\mathrm{(drv)}}=p_cl_c\hbar(\Gamma\rho_{ee}+\dot{\rho}_{ee}).
\end{equation}
Equation \eqref{b12d} shows that the total torque $T_z^{\mathrm{(drv)}}+Q_z^{\mathrm{(drv)}}$ resulting from the interaction with the driving field 
is the product of the Minkowski angular momentum $p_cl_c\hbar$ of a guided light photon and the photon-absorption rate $\Gamma\rho_{ee}+\dot{\rho}_{ee}$.
Thus, the total torque produced by the interaction with the driving field is equal to the angular momentum of the driving photons absorbed per unit of time.

We find that the ratio between the orbital and spin torques $T_z^{\mathrm{(drv)}}$ and $Q_z^{\mathrm{(drv)}}$ produced by the interaction with the driving field is 
\begin{equation}\label{b12c}
\frac{T_z^{\mathrm{(drv)}}}{Q_z^{\mathrm{(drv)}}}=\frac{p_cl_c-q}{q}. 
\end{equation}
This ratio is determined by the azimuthal mode order $l_c$, the mode polarization index $p_c$,  and the dipole polarization index $q$. However, it does not depend on the radial distance $r$ and the fiber radius $a$. 
Meanwhile, the ratio $J_z^{\mathrm{(orb)}}/J_z^{\mathrm{(spin)}}$ between
the orbital part $J_z^{\mathrm{(orb)}}$ and the spin part $J_z^{\mathrm{(spin)}}$ of the axial angular momentum $J_z$ of the guided driving field depends on $a$ (see Appendix \ref{sec:appendix}).
Therefore, we have  $T_z^{\mathrm{(drv)}}/Q_z^{\mathrm{(drv)}}\not=J_z^{\mathrm{(orb)}}/J_z^{\mathrm{(spin)}}$.
This inequality means that the orbital and spin angular momenta of the guided field are not transferred separately to the orbital and spin angular momenta of the two-level atom, unlike the case of small isotropic particles in free space \cite{Dholakia2003}.

We note that, for $l_c\geq1$ and $q=p_c$, we have $(p_cl_c-q)\hbar=p_c(l_c-1)\hbar$. 
In this case, Eq.~\eqref{b7} indicates that the photon angular momentum is converted into the atomic spin and orbital angular momenta.
For $l_c\geq1$ and $q=-p_c$, we have $(p_cl_c-q)\hbar=p_c(l_c+1)\hbar$. In this case, Eq.~\eqref{b7} says that the photon angular momentum and the change of the atomic spin angular momentum have the same sign and add up in generating the atomic orbital angular momentum.
For $l_c\geq1$ and $q=0$, the total photon angular momentum is converted into the atomic orbital angular momentum. 
  
Equation \eqref{b7} can be used for not only hybrid modes ($l_c\ge1$) but also TE and TM modes ($l_c=0$).
In the cases of TE and TM modes, we  have 
\begin{equation}\label{b12h}
T_z^{\mathrm{(drv)}}= -Q_z^{\mathrm{(drv)}} =-q\hbar(\Gamma\rho_{ee}+\dot{\rho}_{ee}), 
\end{equation}
which indicates that
the atomic orbital angular momentum can be generated from the atomic spin angular momentum through the interaction with a photon in a TE or TM mode
having no angular momentum.
This conversion of atomic spin angular momentum into atomic orbital angular momentum via the interaction with a guided photon
is possible because the guided mode is a structured field with a complex polarization profile.
When an atom with a $\pi$, $\sigma_+$, or $\sigma_-$ transition interacts with a plane-wave field in free space, in accordance with the selection rules, the atomic spin angular momentum is converted only to the photon spin angular momentum.

Note that, in the case where the atom is at rest and in the steady-state regime, we have $\dot{\rho}_{ee}=0$. 
In this case, we obtain 
\begin{equation}\label{b12e}
T_z^{\mathrm{(drv)}}=(p_cl_c-q)\hbar\Gamma\rho_{ee} 
\end{equation}
and 
\begin{equation}\label{b12f}
Q_z^{\mathrm{(drv)}}=q\hbar\Gamma\rho_{ee}, 
\end{equation}
where the population $\rho_{ee}$ of the excited state of the atom in the steady state is given as \cite{coolingbook}
\begin{equation}\label{b12g}
\rho_{ee}=\frac{|\Omega|^2}{4\Delta^2+\Gamma^2+2|\Omega|^2}.
\end{equation}
In the limit of high driving field powers, we have $\rho_{ee}\to 1/2$, which leads to the limiting values
\begin{equation}\label{b12i}
\begin{split}
T_z^{\mathrm{(drv)}}&\to (p_cl_c-q)\hbar\Gamma/2,\\ 
Q_z^{\mathrm{(drv)}}&\to q\hbar\Gamma/2.
\end{split}
\end{equation}

\begin{figure}[htbp]
\centering\includegraphics{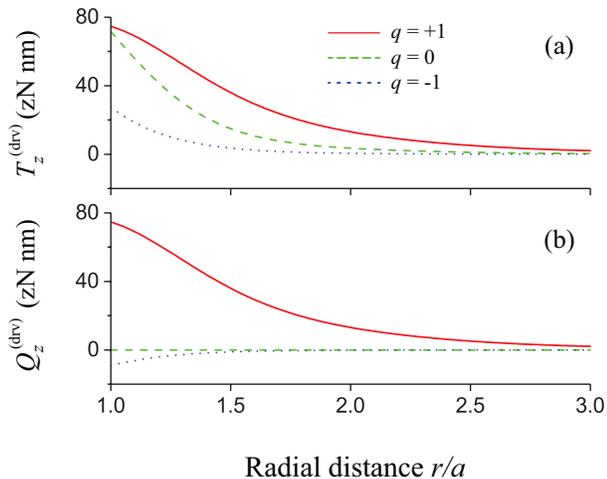}
\caption{Radial dependencies of the orbital and spin torques $T_z^{\mathrm{(drv)}}$ (a) and
$Q_z^{\mathrm{(drv)}}$ (b) of the guided driving field on the atom being at rest and in the steady state. 
The dipole matrix-element vector $\mathbf{d}$ has only one nonzero spherical tensor component $d_q$, where $q=1$, 0, and $-1$.
The driving field is in a quasicircularly polarized HE$_{21}$ mode with $f_c=+1$ and $p_c=+1$. 
The power and detuning of the driving field are chosen to be $P=1$ pW and $\Delta=0$. 
The fiber radius is $a=350$ nm. 
The dipole magnitude $d$ corresponds to the natural linewidth $\gamma_0/2\pi=6.065$ MHz of the $D_2$ line of a $^{87}$Rb atom. 
The wavelength of the atomic transition is $\lambda_0=780$ nm. 
The refractive indices of the fiber and the vacuum cladding are $n_1=1.4537$ and $n_2=1$, respectively.
}
\label{fig2}
\end{figure}

We plot in Fig.~\ref{fig2} the torques $T_z^{\mathrm{(drv)}}$ and $Q_z^{\mathrm{(drv)}}$ as functions of the radial position $r$ of the atom in the case where
the driving field is in a quasicircularly polarized HE$_{21}$ mode and the atom is at rest and in the steady-state regime. The fact that the solid red curves of the figure have the same sign indicates that, for $q=p_c$, the angular momentum of guided light is converted into the orbital and spin angular momenta of the atom in the excitation process. The opposite signs of the dotted blue curves in Figs.~\ref{fig2}(a) and \ref{fig2}(b) indicate that, for $q=-p_c$, the atomic spin angular momentum is converted into the atomic orbital angular momentum due to the excitation of the atom by guided light.

We note that the maximal values of the torques in Fig.~\ref{fig2} are on the order of 80 zN nm. 
For the axial orbital torque on the atom at the radial distance of 400 nm, the corresponding azimuthal force is on the order of $0.2$ zN. Such a value is comparable to the optical forces on single atoms in laser cooling and trapping techniques \cite{coolingbook}. By increasing the power of the guided driving field, we can approach the limiting values \eqref{b12i}.

We now calculate the axial orbital torque $T_z^{\mathrm{(spon)}}$ produced by the recoil of spontaneous emission. 
The expressions for the coupling coefficients $G_{\alpha=\mu,\nu}$ are given by Eqs.~\eqref{c14} and \eqref{c15}.
In the case where a single spherical tensor component $d_q$ of the dipole matrix element vector $\mathbf{d}$ is nonzero, 
we find 
\begin{equation}\label{b15a}
\begin{split}
\frac{G_{\mu}}{\partial\varphi}&=i(pl-q) G_{\mu},\\
\frac{G_{\nu}}{\partial\varphi}&=i(l-q)G_{\nu}.
\end{split}
\end{equation}
In this case, the axial component of the orbital torque of spontaneous emission recoil is found from Eq.~\eqref{c27a} to be
\begin{equation}\label{b15}
T_z^{\mathrm{(spon)}}=q\hbar\Gamma-\hbar\sum_{\mu_0}pl\gamma_{\mu_0}-\hbar\sum_{\nu_0}l\gamma_{\nu_0},
\end{equation}
where 
\begin{equation}\label{b15b}
\begin{split}
\gamma_{\mu_0}&=2\pi|G_{\mu_0}|^2,\\ 
\gamma_{\nu_0}&=2\pi|G_{\nu_0}|^2 
\end{split}
\end{equation}
are the rates of spontaneous emission into the guided mode $\mu_0$ and the radiation mode $\nu_0$ \cite{sponhigh}.

Equation \eqref{b15} describes the exchange of angular momentum between the quantum field and the atom in the spontaneous emission process.
Indeed, the angular momentum of a photon emitted into a guided mode $\mu=(\omega N f p)$ or a radiation mode $\nu=(\omega \beta l p)$ is $pl\hbar$ or $l\hbar$,
respectively, and the change of the spin angular momentum of the atom due to a downward transition is $-q\hbar$. 
Then, it is clear from Eq.~\eqref{b15} that the angular momentum of re-emitted photons is converted into the atomic spin and orbital angular momenta.
Thus, we observe again the conservation of the total angular momentum of the atom-field system and the agreement with the Minkowski formulation of angular momentum of light.
Equation \eqref{b15} is another key result of the present paper.

The axial spin torque $Q_z^{\mathrm{(spon)}}$ produced by the spontaneous emission process is the product of the change $-q\hbar$ of the spin angular momentum of the atom per downward transition and the spontaneous emission rate $\Gamma$, that is,
\begin{equation}\label{b16}
Q_z^{\mathrm{(spon)}}=-q\hbar\Gamma.
\end{equation}
We find 
\begin{equation}\label{b16a}
T_z^{\mathrm{(spon)}}+Q_z^{\mathrm{(spon)}}=-\hbar\sum_{\mu_0}pl\gamma_{\mu_0}-\hbar\sum_{\nu_0}l\gamma_{\nu_0}.
\end{equation}
Equation \eqref{b16a} shows that the total torque $T_z^{\mathrm{(spon)}}+Q_z^{\mathrm{(spon)}}$ of spontaneous emission on the atom
is equal to the minus of the angular momentum of the photons that are spontaneously emitted from the atom per unit of time.

\begin{figure}[htbp]
\centering\includegraphics{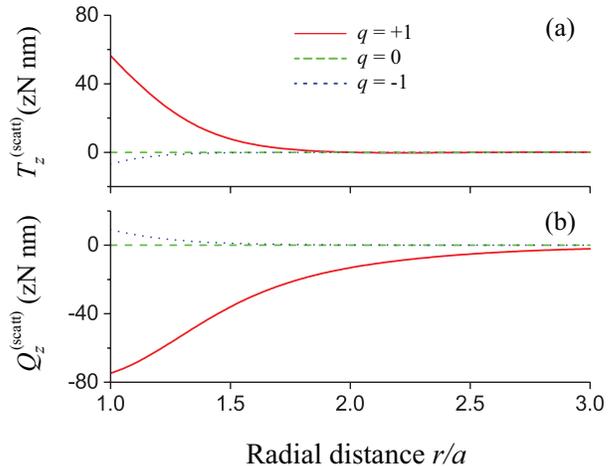}
\caption{Radial dependencies of the orbital and spin scattering torques 
$T_z^{\mathrm{(scatt)}}=\rho_{ee}T_z^{\mathrm{(spon)}}$ and $Q_z^{\mathrm{(scatt)}}=\rho_{ee}Q_z^{\mathrm{(spon)}}$ for the parameters of Fig.~\ref{fig2}.}
\label{fig3}
\end{figure}

It is clear that the torques $T_z^{\mathrm{(spon)}}$ and $Q_z^{\mathrm{(spon)}}$ of spontaneous emission on the atom do not depend on the driving field. However, the scattering torques $T_z^{\mathrm{(scatt)}}=\rho_{ee}T_z^{\mathrm{(spon)}}$ and $Q_z^{\mathrm{(scatt)}}=\rho_{ee}Q_z^{\mathrm{(spon)}}$ depend on the driving field through the excited-state population $\rho_{ee}$.

We plot in Fig.~\ref{fig3} the torques $T_z^{\mathrm{(scatt)}}$ and $Q_z^{\mathrm{(scatt)}}$ as functions of the radial position $r$ of the atom being at rest and in the steady-state regime. 
The opposite signs of the curve for $T_z^{\mathrm{(scatt)}}$ for a given $q\not=0$ and the corresponding curve for $Q_z^{\mathrm{(scatt)}}$ indicate that the atomic spin angular momentum is converted into the atomic orbital angular momentum due to the scattering of light from the atom.

Finally, we discuss the axial torque components  $T_z^{(\mathrm{vdW})e}$ and $T_z^{(\mathrm{vdW})g}$, produced by the van der Waals potentials $U_e$ and $U_g$. 
Depending on the orientation of the dipole matrix-element vector $\mathbf{d}$, 
the  potentials $U_e$ and $U_g$ may depend on the azimuthal angle $\varphi$ and, hence,
the torques $T_z^{(\mathrm{vdW})e}$ and $T_z^{(\mathrm{vdW})g}$  may be nonzero.
In this paper, we consider the case where a single spherical tensor component $d_q$ of the dipole matrix element vector $\mathbf{d}$ is nonzero.  In this case, due to the symmetry of the dipole with respect to the fiber, 
the  potentials $U_e$ and $U_g$  are independent of $\varphi$ and, therefore, we have $T_z^{(\mathrm{vdW})e}=T_z^{(\mathrm{vdW})g}=0$.
 
Combining the above results, we find that the total axial orbital torque is $T_z=T_z^{\mathrm{(drv)}}+\rho_{ee}T_z^{\mathrm{(spon)}}$ and reads
\begin{equation}\label{b17}
T_z=\hbar\rho_{ee}\Big( p_cl_c\Gamma-\sum_{\mu_0}pl\gamma_{\mu_0}-\sum_{\nu_0}l\gamma_{\nu_0}\Big)
+(p_cl_c-q)\hbar \dot{\rho}_{ee}.
\end{equation}
The total axial spin torque is $Q_z=Q_z^{\mathrm{(drv)}}+\rho_{ee}Q_z^{\mathrm{(spon)}}$ and reads
\begin{equation}\label{b18}
Q_z=q\hbar\dot{\rho}_{ee}.
\end{equation}
When the atom is at rest and in the steady-state regime, we have $\dot{\rho}_{ee}=0$. 
In this case, we obtain
\begin{equation}\label{b18a}
T_z=\hbar\rho_{ee}\Big( p_cl_c\Gamma-\sum_{\mu_0}pl\gamma_{\mu_0}-\sum_{\nu_0}l\gamma_{\nu_0}\Big)
\end{equation}
and 
\begin{equation}\label{b18b}
Q_z=0.
\end{equation}

\begin{figure}[htbp]
\centering\includegraphics{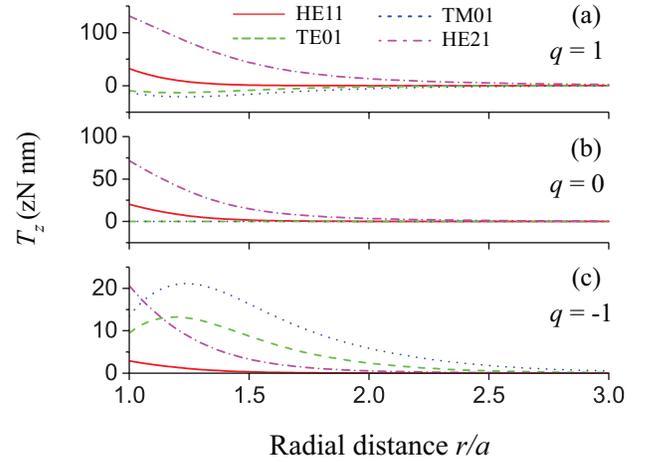}
\caption{Radial dependence of the total orbital torque $T_z$ on the atom being at rest and in the steady state.
The driving field is in a quasicircularly polarized  HE$_{11}$ mode (solid red curves), a TE$_{01}$ mode (dashed green curves), a TM$_{01}$ mode (dotted blue curves), or a quasicircularly polarized  HE$_{21}$ mode (dashed-dotted magenta curves), with the power $P=1$ pW. 
The polarization circulation index for the fields in the HE$_{11}$ and HE$_{21}$ modes is $p_c=+1$. 
Other parameters are as for Fig.~\ref{fig2}.
}
\label{fig4}
\end{figure}

We plot in Fig.~\ref{fig4} the total axial orbital torque $T_z$ as a function of the radial position $r$ of the atom at rest and in the steady-state regime.
The results of calculations for different types of guided modes with a given power are shown. We observe from Fig.~\ref{fig4}(a) that,  for $q=1$, the axial component $T_z$ of the total orbital torque is larger for the HE$_{11}$ and HE$_{21}$ modes with $p_c=q=1$ than for the TE$_{01}$ and TM$_{01}$ modes. However, Fig.~\ref{fig4}(c) shows that,  for $q=-1$, in the region  $r/a>1.15$, $T_z$ is larger for the TE$_{01}$ and TM$_{01}$ modes than for the HE$_{11}$ and HE$_{21}$ modes with $p_c=-q=1$. 
The occurrence of this feature is due to the fact that, for $q=-1$, the Rabi frequency $\Omega$ is larger for the TE$_{01}$ and TM$_{01}$ modes than for the HE$_{11}$ and HE$_{21}$ modes with $p_c=-q=1$.

\section{Summary}
\label{sec:summary}

In conclusion, we have calculated analytically and numerically the axial orbital and spin torques of guided light on a two-level atom near an optical nanofiber. With this we have shown that the generation of these torques is governed by the angular momentum conservation law with the photon angular momentum in the Minkowski formulation.
In addition, we have found that, unlike the orbital torque on an atom in free space, the orbital torque on the atom near the fiber has a contribution  from the average recoil of spontaneously emitted photons. 
We have shown that the photon angular momentum and the atomic spin angular momentum can be converted into the atomic orbital angular momentum. We have found that the orbital and spin angular momenta of the guided field are not transferred separately to the orbital and spin angular momenta of the atom, unlike the case of small isotropic particles in free space. 
Our results  quantify the transfer of angular momentum of guided photons to atoms and are important when trying to generate, control, and manipulate the orbital and spin angular momenta of atoms, molecules, and particles using nanofiber guided light. They can be expected to have significant influence on ongoing and future experiments in nanofiber optics. 

\appendix

\section{Angular momentum of guided light in the Minkowski formulation}
\label{sec:appendix}

For the linear momentum density of the field in a dielectric medium, 
the Abraham formulation takes 
$\mathbf{p}=\mathbf{p}_{\mathrm{A}}\equiv[\mathbf{E}\times \mathbf{H}]/c^2=\mathbf{S}/c^2$, where
$\mathbf{S}=[\mathbf{E}\times \mathbf{H}]$ is the Poynting vector.
On the other hand, the Minkowski formulation takes  
$\mathbf{p}=\mathbf{p}_{\mathrm{M}}\equiv[\mathbf{D}\times \mathbf{B}]=n^2\mathbf{S}/c^2$. 
Angular momentum of guided light has been studied in the Abraham \cite{Partanen2018a,Fam2006,highorder} and Minkowski \cite{Partanen2018a,Fam2006,Bliokh2018} formulations. In this appendix, we present a simple derivation for the angular momentum of guided light in the
Minkowski formulation. 

The Minkowski angular momentum of guided light per unit length is defined by 
\begin{equation}\label{m17}
\mathbf{J}\equiv\int (\mathbf{R}\times\mathbf{p}_{\mathrm{M}})\,d\mathbf{r}
=\frac{1}{c^2}\int n^2 (\mathbf{R}\times\mathbf{S})\,d\mathbf{r}.
\end{equation}
Here, we have introduced the notation $\int d\mathbf{r}=\int_0^{2\pi}d\varphi\int_0^{\infty}r\,dr$ for the integral over the cross-section plane and the notation $\mathbf{R}=x\hat{\mathbf{x}}+y\hat{\mathbf{y}}+z\hat{\mathbf{z}}$ for the position vector in the three-dimensional space. The refractive index $n$ is a function of the radial position $r$ and is given as $n(r)=n_1$ for $r<a$ and $n_2$ for $r>a$.

The Minkowski angular momentum $\mathbf{J}$ of a light beam can be decomposed into orbital and spin parts as
\cite{Bliokh2017a,Bliokh2017b} 
\begin{equation}\label{m19d}
\mathbf{J}=\mathbf{J}^{(\mathrm{orb})}+\mathbf{J}^{(\mathrm{spin})}.
\end{equation}
In the dual-symmetric formalism, the  orbital and spin parts of angular momentum per unit length are given as
\begin{equation}\label{m19c}
\mathbf{J}^{(\mathrm{orb})}=\int \mathbf{j}^{(\mathrm{orb})} \,d\mathbf{r} 
\end{equation}
and 
\begin{equation}\label{m19b}
\mathbf{J}^{(\mathrm{spin})}=\int \mathbf{j}^{(\mathrm{spin})} \,d\mathbf{r}, 
\end{equation}
where
\begin{equation}\label{m19a}
\begin{split}
\mathbf{j}^{(\mathrm{orb})}&=\frac{\epsilon_0}{4\omega} n^2\mathrm{Im}[\boldsymbol{\mathcal{E}}^*\cdot(\mathbf{R}\times\nabla)\boldsymbol{\mathcal{E}}]\\ 
&\quad +\frac{\mu_0}{4\omega} \mathrm{Im}[\boldsymbol{\mathcal{H}}^*\cdot(\mathbf{R}\times\nabla)\boldsymbol{\mathcal{H}}]
\end{split}
\end{equation}
and
\begin{equation}\label{m20a}
\mathbf{j}^{(\mathrm{spin})}=\frac{\epsilon_0}{4\omega} n^2\mathrm{Im}(\boldsymbol{\mathcal{E}}^*\times \boldsymbol{\mathcal{E}}) 
+\frac{\mu_0}{4\omega} \mathrm{Im}(\boldsymbol{\mathcal{H}}^*\times \boldsymbol{\mathcal{H}})
\end{equation}
are the corresponding densities \cite{highorder,Bliokh2018}. Here, $\boldsymbol{\mathcal{E}}$ and $\boldsymbol{\mathcal{H}}$
are the envelopes of the positive frequency components of the electric and magnetic parts $\mathbf{E}$ and $\mathbf{H}$ of the field.
In Eq.~\eqref{m19a}, the dot product applies to the field vectors, that is, 
$\boldsymbol{\mathcal{A}}\cdot(\mathbf{R}\times\boldsymbol{\nabla})\boldsymbol{\mathcal{B}}\equiv \sum_{i=x,y,z}\mathcal{A}_i(\mathbf{R}\times\boldsymbol{\nabla})\mathcal{B}_i$ for arbitrary field vectors $\boldsymbol{\mathcal{A}}$  and $\boldsymbol{\mathcal{B}}$.

We introduce the notation $\mathbf{j}^{(\mathrm{can})}=\mathbf{j}^{(\mathrm{orb})}+\mathbf{j}^{(\mathrm{spin})}$
for the sum density of the orbital and spin parts of angular momentum. 
The quantity $\mathbf{j}^{(\mathrm{can})}$ is called the canonical Minkowski angular momentum density. 
Although $\mathbf{J}=\int \mathbf{j}^{(\mathrm{can})}\,d\mathbf{r}=\int (\mathbf{R}\times\mathbf{p}_{\mathrm{M}})\,d\mathbf{r}$, we have, in general, $\mathbf{j}^{(\mathrm{can})}\not=(\mathbf{R}\times\mathbf{p}_{\mathrm{M}})$ \cite{Bliokh2018}.

Note that TE and TM modes and quasilinearly polarized HE and EH hybrid modes have no angular momentum.
Therefore, we consider below only quasicircularly polarized HE and EH hybrid modes. 

For quasicircularly polarized hybrid modes of the fiber, the full mode functions are given by
\begin{equation}\label{m3}
\begin{split}
\boldsymbol{\mathcal{E}} &=\mathcal{A} (\hat{\mathbf{r}}e_r+p\hat{\boldsymbol{\varphi}}e_\varphi+f\hat{\mathbf{z}}e_z) e^{if\beta z +ipl\varphi},\\
\boldsymbol{\mathcal{H}} &=\mathcal{A} (fp\hat{\mathbf{r}}h_r+f\hat{\boldsymbol{\varphi}}h_\varphi+p\hat{\mathbf{z}}h_z) e^{if\beta z +ipl\varphi},
\end{split}
\end{equation}
where $e_{r,\varphi,z}$ and $h_{r,\varphi,z}$ are the cylindrical components of the reduced electric and magnetic mode functions
$\mathbf{e}$ and $\mathbf{h}$ for the mode with $f=+$ and $p=+$, and $\mathcal{A}$ is a constant determined by the power of the field. The functions $e_{r,\varphi,z}$ and $h_{r,\varphi,z}$ depend on $r$ but not on $\varphi$ and $z$. 
For these modes, the densities 
$j_z^{(\mathrm{orb})}$ and $j_z^{(\mathrm{spin})}$ of the axial components $J_z^{(\mathrm{orb})}=\int j_z^{(\mathrm{orb})} \,d\mathbf{r}$ and 
$J_z^{(\mathrm{spin})}=\int j_z^{(\mathrm{spin})} \,d\mathbf{r}$ of the orbital and spin parts of the angular momentum are found to be
\begin{equation}\label{m22}
\begin{split}
j_z^{(\mathrm{orb})}&=|\mathcal{A}|^2\bigg\{p\frac{\epsilon_0}{4\omega} n^2[l|\mathbf{e}|^2-2\mathrm{Im}(e_r^*e_\varphi)] \\
&\quad+p\frac{\mu_0}{4\omega} [l|\mathbf{h}|^2-2\mathrm{Im}(h_r^*h_\varphi)]\bigg\} \\
\end{split}
\end{equation}
and
\begin{equation}\label{m23}
j_z^{(\mathrm{spin})}=|\mathcal{A}|^2\bigg[p\frac{\epsilon_0}{2\omega} n^2\mathrm{Im}(e_r^*e_\varphi) 
+p\frac{\mu_0}{2\omega}\mathrm{Im}(h_r^*h_\varphi)\bigg].
\end{equation}
We note that very similar expressions have been derived for the Abraham formulation \cite{highorder}.
It follows from Eqs.~\eqref{m22} and \eqref{m23} that the canonical density $j_z^{(\mathrm{can})}=j_z^{(\mathrm{orb})}+j_z^{(\mathrm{spin})}$ of the  axial component $J_z$ of the Minkowski angular momentum is
\begin{equation}\label{m23a}
j_z^{(\mathrm{can})}=|\mathcal{A}|^2\bigg(pl\frac{\epsilon_0}{4\omega} n^2|\mathbf{e}|^2 + pl\frac{\mu_0}{4\omega} |\mathbf{h}|^2\bigg).
\end{equation}
Meanwhile, the energy per unit length is given by 
$U=\int u \,d\mathbf{r}$, where 
\begin{equation}\label{m10}
u=|\mathcal{A}|^2\bigg(\frac{\epsilon_0}{4} n^2|\mathbf{e}|^2+\frac{\mu_0}{4} |\mathbf{h}|^2\bigg)
\end{equation}
is the energy density. Comparison between Eqs.~\eqref{m23a} and \eqref{m10} shows that the angular momentum per photon in the canonical Minkowski formulation is \cite{Partanen2018a,Bliokh2018}
\begin{equation}\label{m23b}
j_z^{(\mathrm{ph})}=\hbar\omega\frac{J_z}{U}=\hbar\omega\frac{j_z^{(\mathrm{can})}}{u}=pl\hbar.
\end{equation}
Thus, the Minkowski angular momentum of a photon in a quasicircularly polarized hybrid guided mode is an integer multiple of $\hbar$ \cite{Partanen2018a,Fam2006,Bliokh2018}.

Equations \eqref{m23a}--\eqref{m23b} show that the ratio $j_z^{(\mathrm{can})}/{u}$ 
between the canonical axial angular momentum density $j_z^{(\mathrm{can})}$  and the energy density $u$
does not depend on the radial distance $r$ and the fiber radius $a$.
In general, the ratio $j_z^{(\mathrm{orb})}/j_z^{(\mathrm{spin})}$ between the
orbital and spin components of $j_z^{(\mathrm{can})}$ is a function of $r$ and $a$,
and the ratio $J_z^{(\mathrm{orb})}/J_z^{(\mathrm{spin})}$ between the
orbital and spin components of $J_z$ is a function of $a$. Our additional numerical calculations, which are not shown here,
confirm these dependencies.

Note that Eq.~(\ref{m23b}) is valid for not only quasicircularly polarized HE and EH hybrid guided modes but also TE and TM guided modes. When we perform similar calculations for radiation modes $\nu=(\omega \beta l p)$, we find a similar result: $j_z^{(\mathrm{ph})}=l\hbar$.

\section*{Acknowledgments}

This work was supported by the Okinawa Institute of Science and Technology Graduate University.


\end{document}